\documentclass[10pt, final,journal,letterpaper]{IEEEtran}
\usepackage{cite}

\usepackage[dvips]{graphicx}

\graphicspath{{../}}
\DeclareGraphicsExtensions{.eps}

\begin{document}
\title{Digital logic using 3-terminal spin transfer torque devices}

\author{\IEEEauthorblockN{Benjamin Buford, Albrecht Jander, and Pallavi Dhagat} 
\IEEEauthorblockA{\\School of Electrical Engineering
and Computer Science\\ Oregon State University\\
Corvallis, OR\\ Email: jander@eecs.oregonstate.edu}}


\markboth{}%
{Buford \MakeLowercase{\textit{et al.}}: Digital logic using 3-terminal spin transfer torque devices}

\maketitle

\begin{abstract}

We demonstrate for the first time that functionally complete digital logic can be created by using three terminal devices each consisting of a magnetic tunnel junction (MTJ) and spin transfer torque (STT) element with a shared free magnetic layer. Rather than the output of devices being high or low voltage levels, logical states are represented as output resistances which retain their state when unpowered. By cascading and clocking devices using a multi-phase clock, threshold logic can be used to implement functionally complete digital logic. Using this approach, a single device can act as an AND, NAND, OR, or NOR gate without the need for external circuitry. Required device parameters and tolerances are discussed.

\end{abstract}


\IEEEpeerreviewmaketitle

\section{Introduction}
\IEEEPARstart{D}{igital} logic has traditionally been implemented using complementary metal-oxide-semiconductor (CMOS) transistors, but these devices are approaching their intrinsic scalability limit as device sizes approach the sub-nanometer scale and off-state leakage current becomes intolerable\cite{continuedTransistorScaling, GedankenModel}. Spintronic devices show promise as an alternative to CMOS technology as most spintronic properties scale inversely with device size.  Some spintronic logic devices, such as spin accumulation magnetologic gates\cite{SemiconductorMagnetologic}, magnetic tunnel junction logic devices\cite{MTJLogic}, and domain wall logic devices \cite{domainWallLogic} have shown feasible approaches to spintronic logic, but rely on inherently inefficient magnetic fields generated by traces carrying high currents. Recently an all-spin logic device, which relies only on spin polarized currents to set the orientation of magnetic layers, has been proposed \cite{ASLD}. This method eliminates the need for Amperian magnetic fields and instead relies on spin transfer torque to set the orientation of magnetic layers.

Spin transfer torque (STT) was first presented in 1994 by Slonczewski \cite{SlonczewskiSTT} as a method of setting the orientation of magnetization in a ferromagnet using a spin polarized current. Spin transfer torque induced switching requires high current densities of up to $10^7$ A/cm$^2$, and has shown a giant magnetoresistive effect on the order of 10\%\cite{cornellThreeTerminal}. The efficiency of spin transfer torque scales inversely with device size. 

Spin transfer torque, combined with the ability of magnetic tunnel junctions (MTJs) to demonstrate the high tunneling magnetoresistance (TMR) effects (of up to 604\% \cite{HighTMR} at room temperature), provides the basis for three terminal devices that can set the orientation of a magnetic free layer using current through one current path, and can read the orientation of the free layer through another\cite{IBMThreeTerminal, cornellThreeTerminal}.

\begin{figure}[!t]
\centering
\includegraphics[width =250pt]{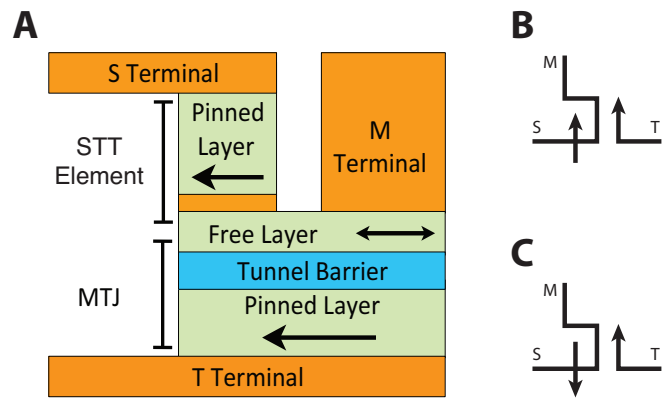}
\caption{Device topology (A) and proposed schematic symbol for non-inverting (B) and inverting (C) devices.}
\label{fig:DeviceDrawingAndSymbols}
\end{figure}

Here, we propose a method of cascading these three terminal devices to implement digital logic. In this paper we will discuss logic circuits created using the three terminal device presented by Braganca et al. in \cite{cornellThreeTerminal}, and shown in Figure~\ref{fig:DeviceDrawingAndSymbols}. This device consists of an MTJ which shares a free layer with a spin transfer torque element, and has a third electrical contact made to the free layer. The approach presented here will work for any three terminal device exhibiting similar terminal behavior.

\section{Principles of operation}

For ease of discussing connectivity of devices we label the three terminals T, M and S, corresponding to the tunnel junction, middle contact, and STT element respectively. We propose connecting devices in a manner seen in Figure~\ref{fig:BufferSchematic}. The T terminal acts as an output which has low or high output resistance corresponding to the orientation of the free layer. Two clock lines are connected to the M terminals of alternating devices, but only one clock line is active at a time while the other serves as a ground connection. When a device is being clocked with a positive voltage, current flows through the STT element from the M terminal to the S terminal and then is divided between a shunt resistor to ground and the T terminal of the previous stage. This  causes the magnitude of the current which flows through the STT element to be dependent on the output resistance of the previous stage. Threshold logic can be implemented if the variation of current corresponding to high and low output of the previous stage is large enough to make the difference between reliably switching and reliably not switching the magnetic orientation of the free layer.

Due to the inherent hysteresis in magnetic systems, considerations must be made to allow for digital logic to be implemented without feedback; that is, when clocked, the output of a device should only depend on the output of the previous stage and not its own previous state. Additionally, the state of the previous device should remain unchanged. To accomplish this we use a two phase clock to first preset, and then to evaluate the state of the device. One clock is first pulsed low to unconditionally set the orientation of the free layer to the anti-parallel state (preset), then pulsed high to conditionally set the orientation to a parallel state only if the previous stage had a low resistance output (evaluate). The magnitude of the preset pulse is larger than that of the evaluate pulse to induce a sufficient switching current regardless of the output resistance of the previous stage. Sequential stages of logic are then clocked in an alternating fashion (Figure~\ref{fig:BufferSchematic} C). Since the switching characteristics of spin transfer torque elements are typically asymmetric, i.e. switching the free layer from anti-parallel to parallel magnetization takes less current than switching from parallel to anti-parallel, the difference in clock voltage magnitude may be unnecessary.

\subsection{Single Input Gates}
The buffer and inverter are topologically identical; an inverter is simply a buffer with the magnetic orientation of one the pinned layers reversed during fabrication. If the MTJ contains the reversed pinned layer it allows anisotropic switching characteristics to behave similarly for both inverting and non-inverting devices. 

\begin{figure}[!t]
\centering
\includegraphics[width =250pt]{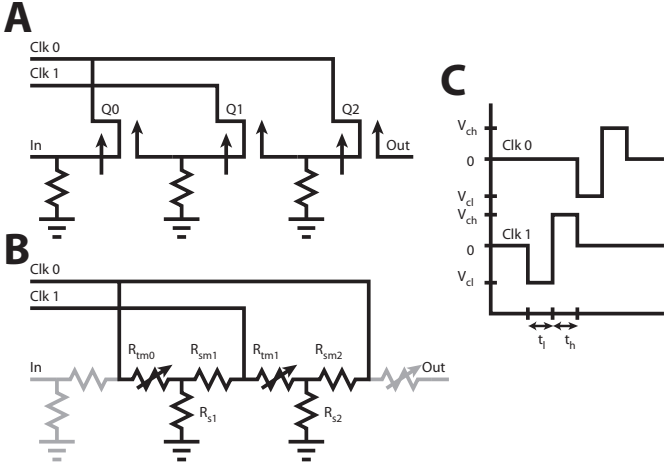}
\caption{A) Three buffers chained together and driven by two clock lines. 
B) The schematic decomposed into junction resistances. Only components involved during the operation of Q1 have been highlighted and labeled. C) Clock timing diagram. Each clock line is first pulsed low during the preset phase, then pulsed high during the evaluate phase. Adjacent stages are clocked alternately to avoid loss of data during the preset during each clock cycle.}
\label{fig:BufferSchematic}
\end{figure}

Consider the buffer Q1 shown in Figure~\ref{fig:BufferSchematic} which can be thought of as two resistors $R_{sm1}$ and $R_{tm1}$ corresponding to the STT element (S to M terminals) and tunnel junction (T to M terminals), where $R_{tm1}$ can take on a high ($R_{AP}$) or low ($R_{P}$) resistance state. The currents $I_{MS}$ and $I_{MT}$, given by (\ref{eq:R_MS_Resistive}) and (\ref{eq:R_MT_Resistive}), are of particular interest, as excessive $I_{MT}$ causes junction breakdown but $I_{MS}$ needs to be sufficiently large to switch the orientation of the free layer. For simplicity, we assume the first order approximation that the free layer magnetization will switch orientation if a sufficient current is applied for a defined pulse time, $t_h$ or $t_l$. This is equivilent to assuming a charge threshold $Q_{th}$ is required for switching.

\begin{equation} \label{eq:R_MS_Resistive}
I_{MS} = \frac{V_{clk}}{R_{sm1} + (R_{tm0} || R_{s1})}
\end{equation}
\begin{equation} \label{eq:R_MT_Resistive}
I_{MT} = \frac{V_{clk}}{R_{tm1} + (R_{sm2} || R_{s2})}
\end{equation}

During the preset pulse the resistance $R_{s1}$ and voltage $V_{ch}$ must have been chosen to meet two conditions. First, the preset pulse must cause a large enough charge to pass through the STT element ($R_{sm1}$) to unconditionally set $R_{tm1}$ to a high state. Second, the charge passing through $R_{sm2}$ must be low enough to not change its state.

During the evaluate phase, switching must only occur when the output of the previous stage is a high resistance. Therefore, the evaluate pulse must either be a lower amplitude or a shorter duration than the preset pulse. This guarantees that as long as adjacent devices were not altered during the preset phase, it will not be altered during the evaluate phase. Additionally $|I_{MS}*t_l|$ must be below $Q_{th}$ when $R_{tm0}$ is in its high resistance state and above $Q_{th}$ when $R_{tm0}$ is in its low resistance state.  

When considering the worst case device state, these conditions yield the following four inequalities that must be met for proper device operation:

\begin{equation} \label{eq:part1}
	\frac{V_{cl}*t_l}{R_{sm1} + (R_{AP} || R_{s1})} > Q_{th}
\end{equation}

\begin{equation} \label{eq:part2}
	\frac{|V_{ch}|*t_h}{R_{sm1} + (R_{AP} || R_{s1})} < Q_{th}
\end{equation}

\begin{equation} \label{eq:part3}
	\frac{|V_{ch}|*t_h}{R_{sm1} + (R_{P} || R_{s1})} > Q_{th}
\end{equation}

\begin{equation} \label{eq:part4}
	\frac{V_{cl}*R_{s2}*t_l}{R_{P}*R_{s2} + R_{P}*R_{sm2} + R_{s2} * R_{sm2}} < Q_{th}
\end{equation}

If these conditions are met, the preset pulse will set the output of Q1 to a high resistance, and if and only if the output of Q0 is a low resistance the evaluate pulse will set the output of Q1 to a low resistance.

\subsection{Multiple Input Gates}

\begin{figure}[!t]
\centering
\includegraphics[width =240pt]{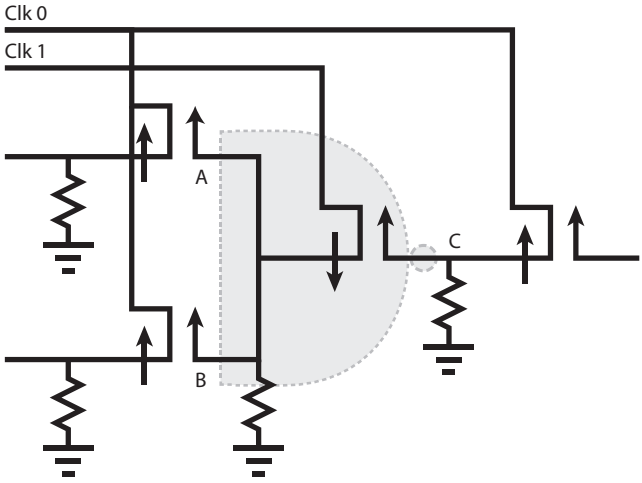}
\caption{Schematic for a NAND gate with inputs A and B. Three peripheral buffers have been added to show connectivity to adjacent stages.}
\label{fig:NANDSchematic}
\end{figure}

Like the buffer and inverter, an AND and NAND gate can each be implemented using a single device and differ from each other only by the orientation of a pinned layer. Since the device state is determined by output resistance of the previous stage, threshold logic can be implemented by putting multiple input devices in parallel, as shown in Figure~\ref{fig:NANDSchematic}. 

\begin{figure}[!t]
\centering
\includegraphics[width =240pt]{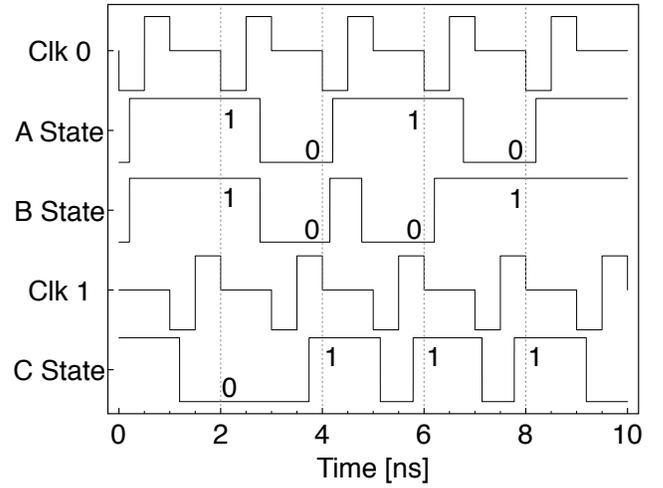}
\caption{Simulated timing diagram for a NAND gate using idealized devices which have an output that instantly switches resistances after current through the STT element exceeds a defined threshold for a defined period of time. A free layer orientation creating a low output resistance is arbitrarily represented as a logical 0, and a high output resistance corresponds to a logical 1.}
\label{fig:NANDTimingDiagram}
\end{figure}

By adjusting the resistance of the attached shunt resistor the resistance threshold can be modified to allow the device to switch during the negative clock pulse when either or both of the inputs are low resistance. A simulated timing diagram of a NAND gate is shown in Figure~\ref{fig:NANDTimingDiagram}. Consider the periods of time where Clk 1 is active; During the preset phase at t=1 ns, Clk 1 is pulled low causing a sufficiently large $I_{MS}$ in the STT element of the NAND gate to cause the free layer to switch to a state anti-parallel to the pinned layer of the STT element. Since this is an inverting device, the free layer is now parallel to the pinned layer of the MTJ and the output is low resistance, or logical 0. Next, during the evaluate phase, Clk 1 is pulled positive, but to a lesser voltage magnitude. Since the resistance of both devices in the previous stage is high, $I_{MS}$ is smaller and insufficient to switch the state back to a high resistance, leaving the output as a logical 0 and ready to be read by the next gate in series. At t=3 ns the device is already in a low resistance state, so the preset pulse does not change the state of the device. The evaluate pulse at t=3.5 ns creates a current large enough to toggle the device as the output resistances of both devices in the previous stage are low.

An OR or NOR gate can be created by adjusting the shunt resistor, effectively shifting the current bias point to cause the evaluate phase to switch switch device state only when both inputs have a low resistance. Threshold logic can be implemented by varying the device size to scale output resistances.

\subsection{Fanout}

As the output state of a device is expressed as an resistance, fanout is only limited by the amount of current the MTJ can sink without damaging it. Having more than one gate connected to the output of a single device requires that the device must sink current directly proportional to the fanout. Since all of the devices in the fanout must be clocked synchronously, the shunt resistor must be decreased to allow a much larger bias which is again proportional to the amount of fanout. If a large fanout is required, device size can be increased to lower the MTJ resistance while keeping the same TMR value. This increases the current the MTJ is able to sink without damage.

\section{Device Paramater Scaling and Tolerance}
One advantage of magnetic based logic over traditional CMOS logic is that energy requirements scale with device size; Since device switching is dependent on a current exceeding a critical current density threshold, the required switching current is proportional to the STT element junction area. Similarly, the applied voltage required to achieve the switching current density is proportional to the resistance of the junctions, and thus scales inversely with device size. 

For use in digital logic device quality can be characterized to first order by the switching current density threshold, the junction area for both the tunnel junction and spin transfer torque element, the TMR ratio, and the resistance area product (RAP) of the tunnel junction when it is in a low resistance state. While most of these device parameters are fabrication process dependent and currently cannot be controlled to great precision, overall logic operation can be designed to work for a wide tolerance of device characteristics by carefully selecting the shunt resistor value as well as the amplitude of the voltage pulses on the clock lines in order to expand the bounds on the switching current given in inequalities~\ref{eq:part1}-\ref{eq:part4}. The following section will be a discussion on how these parameters should be optimized to increase the tolerance to variations in parameters while decreasing overall power usage for the discussed magnetic logic implementation.

\begin{figure}[!t]
\centering
\includegraphics[width=240pt]{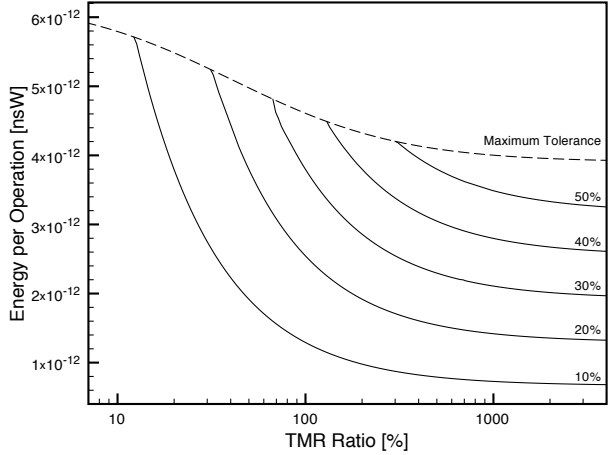}
\caption{Energy per operation for various MTJ TMR ratios where the applied clock voltage and shunt resistor value are selected for various allowable variations in switching current thresholds.}
\label{fig:Energy_vs_TMR}
\end{figure}
To analyze the effects various device parameters have on logic operation we first assume all parameters are equal to those presented in~\cite{cornellThreeTerminal}, then sweep a single device parameter and select an optimal shunt resistor ($R_s$) and clock voltages to maximize the difference in bounds on the switching current in terms of percent change (range over minimum). The required clock voltage can vary from 20 mV up to 5V depending on the required switching current and junction RAP.

Energy per operation scales quadratically with process determined switching current, linearly with the MTJ RAP and device size, and decreases linearly proportional to the STT element RAP. The effect of TMR ratio behaves non-linearly, and is shown in Figure~\ref{fig:Energy_vs_TMR}. The dotted line shows the energy required assuming shunt resistor and voltage amplitude are selected to allow for maximum process variation. If the process can guarantee parameters lie within tighter bounds, optimizations to the shunt resistance can be made to decrease overall power consumption.

\begin{figure}[!t]
\centering
\includegraphics[width =240pt]{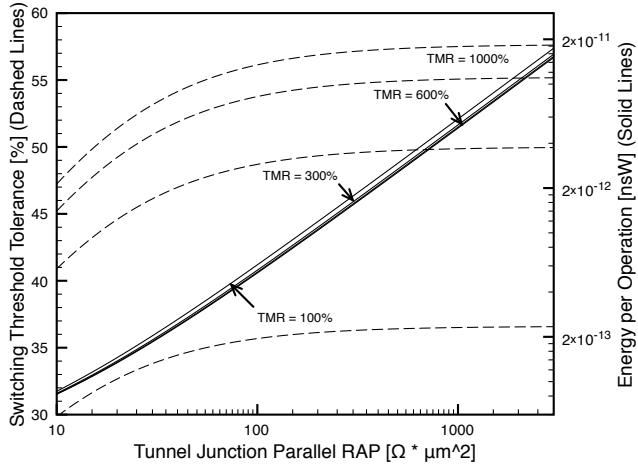}
\caption{Energy per operation and fabrication tolerance vs MTJ RAP for various TMR values.}
\label{fig:TMR_RAP_Sweep}
\end{figure}

The RAP of the MTJ places a limit on both the energy per operation and tolerance to process variation; while a small MTJ RAP significantly reduces energy per operation it causes a decrease in process variation tolerance, as seen in Figure~\ref{fig:TMR_RAP_Sweep}. This suggests that an ideal MTJ RAP is on the order of 100 $\Omega\mu m^2$, but the exact value is dependent on other device properties.

\section{Conclusion}
We have presented a method for implementing arbitrary digital logic using a three terminal device consisting of back-to-back STT junction and MTJ. The implementation requires two out of phase and multi-level clock lines which supply all power to the devices. Digital information is stored as solid-state magnetization of magnetic layers, and is read out as variable conductances when succeeding devices are clocked. To improve the efficiency of these devices device fabrication must be perfected to create three terminal devices with low junction resistance but high TMR ratios and well defined switching current thresholds. As fabrication technology improves, this approach to implementing digital logic has the opportunity to provide low power, high speed, and non-volatile logic operations.

\ifCLASSOPTIONcaptionsoff
  \newpage
\fi


\bibliographystyle{IEEEtran}
\bibliography{IEEEabrv,2010MagneticsPaper}

\end{document}